\documentclass[twoside]{article}
\usepackage{fleqn,espcrc2}
\usepackage{graphicx}
\usepackage{multirow}

\def\simlt{\rlap{\lower 3.5 pt\hbox{$\mathchar \sim$}}\raise 1pt \hbox {$<$}}
\def\simgt{\rlap{\lower 3.5 pt\hbox{$\mathchar \sim$}}\raise 1pt \hbox {$>$}}

\def\lsim{\mathrel{\rlap{\lower3pt\hbox{\hskip0.5pt$\sim$}}\raise1pt\hbox{$<$}}}
\def\gsim{\mathrel{\rlap{\lower3pt\hbox{\hskip0.5pt$\sim$}}\raise1pt\hbox{$>$}}}             
\newcommand{\figscale}{0.4}

\title{
Hadron Spectrum from Dynamical Lattice QCD Simulations
}

\newcommand{\Hiroshima}%
{Department of Physics, Hiroshima University, Higashi-Hiroshima, Hiroshima 739-8526, Japan}

\author{K-I.~Ishikawa\rlap,\address{\Hiroshima}}

\begin{document}
\begin{abstract}
Recent progress in unquenched lattice QCD simulations is reviewed 
with emphasis on understanding of chiral behavior for light quark masses.
\end{abstract}

\maketitle

\section{Introduction}
\label{sec:intro}
In the long standing challenge of the calculation of hadron
spectrum from the first-principles lattice QCD simulation,
the inclusion of dynamical quarks, or unquenching, has been the
crucial but computationally demanding step.
Thanks to the rapid increase of the computer power and
recent developments in the lattice formulations and
numerical algorithms, such simulations have become feasible
in the past several years. 

One of the stumbling block toward full QCD simulations has been 
the treatment of the third dynamical quark, {\it i.e.},  the strange quark.
Since the standard HMC algorithm~\cite{Duane:1987de} assumes 
that the number of
flavor is even, previous simulations including the odd
number of flavors had to use a non-exact algorithm, such as
the $R$ algorithm~\cite{Gottlieb:1987mq}, 
which contains a systematic error of order $\delta\tau^2$, 
with $\delta\tau$ the molecular dynamics step size.
Recently an exact HMC algorithm with polynomial approximation has
been developed~\cite{Takaishi:2001um}
and this difficulty has been essentially removed.

A major problem in the dynamical fermion
simulations is that the up and down quark masses are much
smaller than the masses we can simulate on the lattice.
Therefore, we usually extrapolate the data obtained at
an order of magnitude larger sea quark masses to the
physical point assuming some fitting ansatz. 
For the pseudo-scaler meson channel, the fitting ansatz can
be justified by Chiral Perturbation Theory (ChPT),
which provides a systematic expansion in terms of pion mass
and momentum squared.
However, the problem is that the region of applicability of
ChPT is not known {\it a priori},
and the lattice data, for instance obtained by the JLQCD collaboration,
do not indicate the non-analytic behavior
expected from the next-to-leading order ChPT 
within simulated quark masses, 
{\it e.g., } above $m_s/2$ for JLQCD~\cite{Hashimoto:2002vi}.

Since the last year's Symposium, a number of unquenched lattice QCD
simulations have been pursued to overcome the
problems related to the chiral behavior of hadron spectrum,   
which I review in this talk.
In Section~\ref{sec:unquenched_summary} I summarize
the major unquenched simulations available to date for both
2- and 2+1-flavor cases.
In Section~\ref{sec:2+1} I discuss the algorithmic issues and 
physics results from recent 2+1-flavor simulations.
Discussion on the chiral extrapolation, which is the main focus of 
this review, is given in Section~\ref{sec:chiral_extrap}. 

The theoretical background of the chiral perturbation theory
is discussed by B\"ar\cite{ChPTBaer,ChPTFinVCOL} at this
conference. 
For the unquenched calculations of quark masses, weak matrix
elements, and heavy quark physics, see other reviews in
\cite{WeakLub,FlvWin,QMASRak}.  

\section{Recent Unquenched QCD simulations}
\label{sec:unquenched_summary}

\begin{table*}[tb]
  \centering
  \caption{Recent spectrum runs in dynamical QCD simulations.
($^1$ an update of \cite{AliKhan:2000mw,AliKhan:2001tx}, $^2$ an update of \cite{Allton:1998gi})}
  \begin{tabular}{cccccccc}\hline
    Collab. & $N_f$& Action& $a$ [fm]  & $aL$ [fm]& $M_{\pi}/M_{\rho}$     & $M_{\pi}$ [GeV] & Ref. \\\hline
            &&&&&0.58--0.30 &&\\
\raisebox{1.6ex}[0pt][0pt]{MILC}       & 
\raisebox{1.6ex}[0pt][0pt]{2+1} &
\raisebox{1.6ex}[0pt][0pt]{ SZAT}  &
\raisebox{1.6ex}[0pt][0pt]{0.125, 0.09}&
\raisebox{1.6ex}[0pt][0pt]{ 2.5--3.0} & 
\raisebox{0.1ex}[0pt][0pt]{0.49, 0.38} &
\raisebox{1.6ex}[0pt][0pt]{0.60--0.25}      & 
\raisebox{1.6ex}[0pt][0pt]{\cite{MILCS,Gottlieb:2003bt,MILCNEW}} \\
{\small CP-PACS}  &&&&&&&\\
\raisebox{0.2ex}[0pt][0pt]{{\small \& JLQCD}}    &
             \raisebox{1.6ex}[0pt][0pt]{2+1} &
             \raisebox{1.6ex}[0pt][0pt]{RC(N)} &
             \raisebox{1.6ex}[0pt][0pt]{0.1}  & 
             \raisebox{1.6ex}[0pt][0pt]{1.6, 2.0} &
             \raisebox{1.6ex}[0pt][0pt]{0.77--0.64} &
             \raisebox{1.6ex}[0pt][0pt]{1.00--0.64} &
             \raisebox{1.6ex}[0pt][0pt]{\cite{Kaneko:2003re,NF3L04}} \\\hline
CP-PACS     & 2   & RC(T) & 0.22--0.09& 2.6--2.1 & 0.8--0.5              & 1.2--0.5        & 
\cite{AliKhan:2000mw,AliKhan:2001tx} \\
CP-PACS     & 2   & RC(T) & 0.22      & 2.6      & 0.61--0.35            & 0.58--0.27      & 
\cite{Namekawa:2004bi}\footnotemark[1]\\
UKQCD       & 2   & PC(N) & 0.1       &      1.6 & 0.84--0.58            &  1.1--0.6       & 
\cite{Allton:1998gi}\\
UKQCD       & 2   & PC(N) & 0.1       &      1.6 & 0.44                  & 0.42            & 
\cite{Allton:2004qq}\footnotemark[2]\\
    JLQCD   & 2   & PC(N) & 0.1       & 1.1--1.8 & 0.80--0.60            & 1.37--0.60      &
\cite{Aoki:2002uc}\\
{\small QCDSF} &&&&&&&\\
\raisebox{0.2ex}[0pt][0pt]{{\small \& UKQCD}}  &
              \raisebox{1.6ex}[0pt][0pt]{2}   &
              \raisebox{1.6ex}[0pt][0pt]{PC(N)} &
              \raisebox{1.6ex}[0pt][0pt]{0.1}   &
              \raisebox{1.6ex}[0pt][0pt]{1.2--2.4} &
              \raisebox{1.6ex}[0pt][0pt]{-}       &
              \raisebox{1.6ex}[0pt][0pt]{1.2--0.64} & 
              \raisebox{1.6ex}[0pt][0pt]{\cite{AliKhan:2003cu}} \\
     RBC    & 2   & {\small DBW2DW}& 0.12      &      1.9 & -            & 0.49,0.61,0.70  & 
\cite{RBCNEW} \\
qq+q        & 2   & PW    & 0.20      &      3.2 & -                     & 0.66--0.37      & 
\cite{Farchioni:2003nf,Farchioni:2004tv} \\
     SPQcdR & 2   & PW    & 0.066     & 1.1,1.6  & 0.6--0.8              & - &
\cite{SPQcdR} \\
    GRAL    & 2   & PW    & 0.13, 0.08& 1.1--2.1 & -                     & 0.42--0.64      & 
\cite{Orth:2003nb}\\
SESAM       & 2   & PW    & 0.086     & 1.4      & 0.83--0.69            & 1.00--0.64      & 
\cite{Eicker:1998sy}\\
{\small SESAM } &&&&&&&\\
\raisebox{0.2ex}[0pt][0pt]{{\small \& T$\chi$L}} &
              \raisebox{1.6ex}[0pt][0pt]{2}    &
              \raisebox{1.6ex}[0pt][0pt]{ PW}  & 
              \raisebox{1.6ex}[0pt][0pt]{0.092, 0.076} &
              \raisebox{1.6ex}[0pt][0pt]{1.3--1.8}    &
              \raisebox{1.6ex}[0pt][0pt]{0.83--0.57}  &
              \raisebox{1.6ex}[0pt][0pt]{0.90--0.49}  &
              \raisebox{1.6ex}[0pt][0pt]{\cite{Lippert:1997vg,Eicker:2001dn}}\\\hline
  \end{tabular}
  \label{tab:DYNSIM}
\end{table*}

Table~\ref{tab:DYNSIM} summarizes the recent large-scale
unquenched simulations with $N_f$ dynamical flavors.
The lattice spacing $a$, physical spatial extent of the
lattice $aL$, 
and pseudo-scalar and vector meson masses $M_{PS}$ and
$M_{V}$ at the unitary point 
({\it i.e.,} valence quark mass equals sea quark mass) are listed.

Throughout this paper we use the following abbreviations to 
denote the lattice actions.
The gauge and quark action combinations are denoted as,
PW: plaquette gauge and unimproved Wilson quark actions,
PC(X): plaquette gauge and clover quark actions with the 
improvement coefficient $c_{\mathrm{SW}}$ determined by a
method X,
RC(X): RG-improved gauge and clover quark actions with the 
improvement coefficient determined by a method X,
SZAT : Symanzik improved gauge and AsqTad KS quark actions
$O(a^2)$-improved at tree level,
DBW2DW : DBW2 improved gauge and domain wall quark actions.
The method X for clover quark action is either N:
non-perturbatively determined, or T: tadpole estimated.

Let us touch upon the representative full QCD simulations. 
For the Wilson-type quark action,  the continuum extrapolation
was previously attempted by the CP-PACS
collaboration~\cite{AliKhan:2000mw,AliKhan:2001tx} 
using data at three lattice spacings with $N_f=2$ RC(T)
action, and a better agreement of hadron masses than in the quenched case 
was observed in the continuum limit.
Ref.~\cite{Namekawa:2004bi} is an update
of this work, which extends 
the simulation at the coarsest lattice spacing 
($a\sim 0.22$~fm) toward smaller sea quark masses
corresponding to
$M_{\mathrm{PS}}/M_{\mathrm{V}}=$0.60--0.35.

UKQCD~\cite{Allton:2004qq} also recently extended their previous 
simulation~\cite{Allton:1998gi} toward smaller quark masses.
The SPQcdR collaboration reported their preliminary results
for the light hadron spectrum, light quark masses and
renormalization constants~\cite{SPQcdR}. 
The GRAL collaboration~\cite{Orth:2003nb} has started a study
of the finite volume effect at small quark masses 
on small to medium-sized lattices. All these are $N_f=2$ simulations. 

The CP-PACS and JLQCD collaborations have jointly started $N_f=2+1$ flavor
simulations with the RC(N) action,  and their first results for a 
$16^4\times 32$ lattice appeared at the last lattice
conference~\cite{Kaneko:2003re}. 
They have extend the physical volume to $20^3\times 40$ this year. 

The $N_f=2$ and $2+1$ unquenched simulations with the
KS-type fermions can be performed at smaller quark masses
compared to the Wilson-type quark actions, because the
required computational cost is much reduced.
The 2+1-flavor simulations with 
the SZAT action have been carried out for several years by the MILC
collaboration~\cite{MILCS,Gottlieb:2003bt,MILCNEW}.
This year they reported a detailed analysis of the chiral
and continuum extrapolations by adding data at smaller sea
quark masses and finer lattice spacing. 

With the configuration generated by the MILC collaboration, 
the ``gold-plated'' hadronic observables have been
calculated by the HPQCD-MILC-UKQCD-Fermilab collaborations
and good agreement with experimental results is
observed~\cite{Davies:2003ik}. 
 
The RBC collaboration reported the calculation of
pseudo-scalar meson masses and decay constants using the
unquenched domain wall fermion~\cite{Izubuchi:2003rp,RBCNEW} for 
$N_f=2$ case.

\section{$N_f=2+1$ simulations}
\label{sec:2+1}

\subsection{Algorithmic issues}
The simulation algorithm widely used for unquenched
simulations is the Hybrid Monte Carlo
algorithm~\cite{Duane:1987de}.  
Quarks are treated by the pseudo-fermion
method~\cite{Weingarten:1980hx}, with which we can treat
even number of them dynamically. 
If the single-flavor lattice Dirac operator $D$ has a real
positive determinant $\det[D]$, it is also possible  to
treat it as a probability distribution function for link
variables. 
The naive application of the pseudo-fermion method to
$\det[D]$ results in a complex effective action, however.
This problem is avoidable by constructing an operator $S$
which satisfies $S^2=D$. 
The polynomial approximation and rational approximation can
be used to construct such an
operator~\cite{Lippert:1999up}. 
With this method the HMC algorithm with the polynomial or
rational approximated pseudo-fermion has been obtained.
The algorithm is then combined with the usual two-flavor
HMC algorithm to make an $N_f=2+1$ flavor HMC
algorithm~\cite{Takaishi:2001um}.  
The efficiency of the algorithm is comparable to that of the
HMC algorithm~\cite{Aoki:2001pt}. 

For the KS-type quark actions, the fourth-root trick is 
widely used to express a single flavor of fermion. 
This amounts to taking a fourth-root of the determinant of the KS Dirac
operator $D_{\mathrm{KS}}$, which represents four flavors of
fermions in the continuum.
This trick is combined with the $R$ algorithm~\cite{Gottlieb:1987mq} 
for carrying out simulations.   

An important issue with the $R$ algorithm is controlling 
the systematic error of order
$\delta\tau^2$ at finite molecular dynamics step size
$\delta\tau$.  
It is argued that a condition $\delta\tau < m$ is 
needed~\cite{Clark:2002vz} to avoid large systematic errors. 
A reliable estimate of the magnitude of error is difficult prior to 
the actual simulations.

In the MILC simulations~\cite{MILCS,MILCNEW} 
$\delta \tau \lsim 3/2 am$ is adopted with light dynamical quark  mass $am$.
The systematic error for some observables is investigated 
by comparing the results from additional simulations with 
larger $\delta \tau$. The additional runs are too short
to conclude the $(\delta\tau)^2$ behavior for hadronic masses.
Further investigation on this issue should be made 
by increasing the statistics or using exact algorithms as described below.

The polynomial approximation described for the Wilson-type
quarks above can also be applied to $D_{\mathrm{KS}}^{1/4}$. 
This leads to the exact algorithms, the polynomial HMC which uses
polynomial approximant and the rational HMC which uses rational
approximant, have been developed for the two-flavor
case~\cite{Clark:2003na,Clark:2002vz,Kennedy:1998cu,Aoki:2002xi}.
For more details of the recent development of the simulation
algorithm, see~\cite{Kennedy}.

In order to justify the fourth-root trick one has to show
the existence of a local fermion kernel $D$ which satisfies 
$\det[D_{\mathrm{KS}}]^{1/4}=\det[D]$.  Otherwise there is no
guarantee that the continuum limit is real QCD.
It is shown that a naive candidate $D=D_{\mathrm{KS}}^{1/4}$
is non-local~\cite{Bunk:2004br,Hart:2004sz}, and the
existence of a local operator $D$ is still an open question.
The related issues are discussed
in~\cite{LJansen,Adams,Follana}.

\subsection{$N_f=2+1$ simulation with the Wilson-type fermions}

The CP-PACS and JLQCD collaborations have started a joint
program to perform realistic simulations including the
dynamical strange quark. 
They performed simulations on two lattice volumes,
$16^3\times 32$ and $20^3\times 40$,  at a finite lattice
spacing $a\sim 0.1$~fm with the RG-improved gauge action 
and non-perturbatively $O(a)$-improved Wilson action.
In order to interpolate or extrapolate to the physical
strange quark mass, the simulations are made at two strange
quark masses corresponding to
$M_{\mathrm{PS,SS}}/M_{\mathrm{V,SS}}\simeq$ 0.71--0.77.
For the light up and down quarks which are assumed degenerate, 
six (five) quark masses on 
the $16^3\times 32$ ($20^3\times 40$) lattice are simulated.
It covers the range 
$M_{\mathrm{PS,LL}}/M_{\mathrm{V,LL}}\simeq$ 0.62--0.78.

\renewcommand{\figscale}{0.31}
\begin{figure}[htbp]
  \centering
  \includegraphics[angle=-90,scale=\figscale,clip]{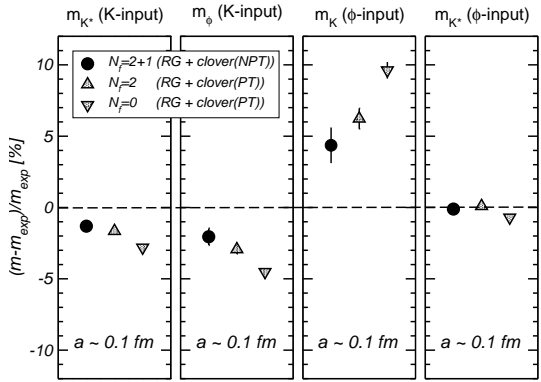}
\vspace*{-2em}
\caption{
  CP-PACS-JLQCD result for the 
  $N_f$ dependence of the light meson masses at $a\sim 0.1$ fm~\cite{NF3L04}.
  Circles represent the $N_f=2+1$ result with
  RC(NP)~\cite{NF3L04}, while 
  triangles show the $N_f=2$ and $0$ data with
  RC(TP)~\cite{AliKhan:2001tx}.
}
\vspace*{-2em}
\label{fig:MESONNf3JLCP}
\end{figure}

\renewcommand{\figscale}{0.32}
\begin{figure*}[htbp]
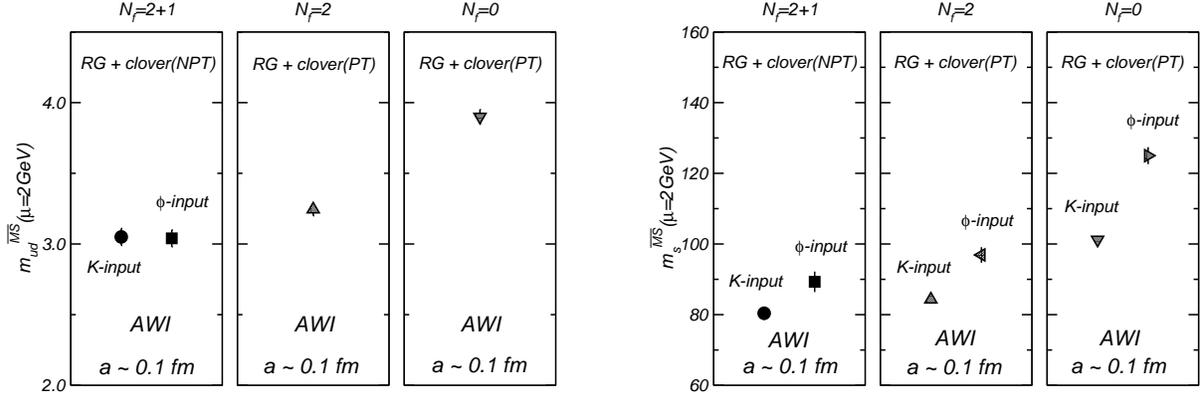

  \centering
  \includegraphics[scale=\figscale,clip]{Figs/Mud_vs_Nf.eps}
  \hfill
  \includegraphics[scale=\figscale,clip]{Figs/Ms_vs_Nf.eps}
\vspace*{-2em}
  \caption{
    $N_f$ dependence of light (left panel) and strange
    (right panel) quark masses at $a\sim 0.1$ fm
    from CP-PACS and JLQCD~\cite{NF3L04}.
}
\vspace*{-1em}
\label{fig:QMASSNf3JLCP}
\end{figure*}

They calculated the light meson mass spectrum, and up and down, 
and strange quark masses. 
The meson masses are calculated at the unitary point where
the valence and sea quarks have equal masses.
Light and strange mesons are constructed with Light-Light
(LL), Light-Strange (LS) and Strange-Strange (SS)
combinations of simulated quarks.
Chiral extrapolation is made by a simultaneous fit 
to LL, LS and SS masses with a polynomial fit function
including up to quadratic terms of valence and sea quark
masses.

Figure~\ref{fig:MESONNf3JLCP} shows the deviation of meson
spectrum from the experiment value.
Lattice data are shown for the 2+1 (filled circle), 2 (up
triangle), and 0 (down triangle) dynamical flavors at $a\sim
0.1$~fm. 
The error bar shows statistical error only.
Compared to the 2-flavor result, the 2+1-flavor simulation
gives the result slightly closer to the experiment.
Figure~\ref{fig:QMASSNf3JLCP} shows the comparison of quark
masses with different $N_f$ at $a\sim 0.1$ fm. 
The reduction of quark mass with the inclusion of dynamical up and 
down quarks is significant. 
The effect of dynamical strange quark is, on the other hand,
not conclusive at this stage, because there may be an effect
of a different choice of $c_{\mathrm{SW}}$ yielding
different scaling violation among the data.
The light quark mass could be sensitive to the choice of the 
detail of the chiral extrapolation.
The effect of the chiral logarithm should be investigated.

\subsection{$N_f=2+1$ simulation with the KS-type fermions}
\label{subsec:KSLH}

The MILC collaboration has been generating configurations
of $N_f=2+1$ QCD with the SZAT action. 
The HPQCD-MILC-UKQCD-Fermilab collaborations calculated the ``gold-plated''
observables on these configurations, and found results shown in 
Fig.~\ref{fig:Nf3GP}.   
A good agreement is seen between the experimental and lattice results 
for $N_f=3$ for a number of quantities covering both 
the light sector and the heavy sector. 

\renewcommand{\figscale}{0.86}
\begin{figure}[tbp]
  \centering
  \includegraphics[scale=\figscale,clip]{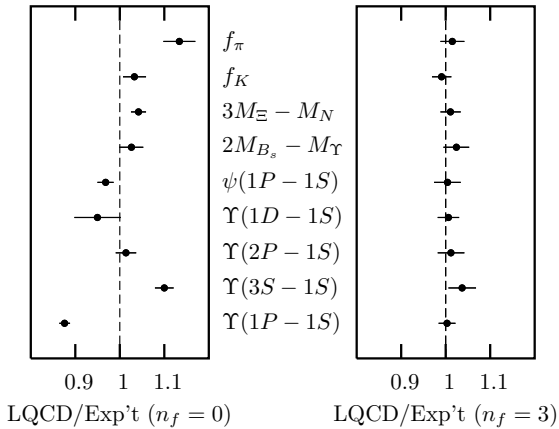}
\vspace*{-4em}
\caption{Lattice QCD results with SZAT action divided by experimental results
for the ``gold-plated'' observables~\cite{Davies:2003ik}. }
\vspace*{-2em}
\label{fig:Nf3GP}
\end{figure}

This year the MILC collaboration published a detailed
analysis of the continuum and chiral
extrapolations~\cite{MILCCHIRAL,MILCNEW}.
Since the analysis of the light pseudo-scalar meson channel
using ChPT is very important, and also instructive to the Wilson-type quark
action simulations, I discuss the key points of
their method to obtain the chiral and continuum limit in Sec.~4.

From this analysis, they obtain the light meson spectrum, the decay constants, 
and the low energy constants of chiral perturbation theory~\cite{MILCCHIRAL}.
The light quark masses are also extracted~\cite{Aubin:2004ck}.
The results for the  pion and $K$ meson decay constants are
\begin{eqnarray}
  f_{\pi}&=&129.5\pm 0.9\pm 3.5 \mbox{\ \ \ MeV},\nonumber\\
  f_{K}  &=&156.6\pm 1.0\pm 3.6 \mbox{\ \ \ MeV},\\
  f_{K}/f_{\pi}&=&1.210(4)(13),\nonumber
\end{eqnarray}
which are updated values of Fig.~\ref{fig:Nf3GP}. The errors are
statistical and systematic, the latter containing  
chiral and continuum extrapolations, the scale determination, and 
electromagnetic effects as carefully investigated in~\cite{MILCNEW}.

The strange quark mass $m_{s}$,
and averaged up and down quark mass $\hat{m}$ in the 
$\mathrm{\overline{MS}}$ scheme at $O(\alpha_s)$ matching are given by 
\begin{eqnarray}
    m_{s}^{\mathrm{\overline{MS}}}&=&76(0)(3)(7)(0)\mbox{\ \ MeV},\nonumber\\
  \hat{m}^{\mathrm{\overline{MS}}}&=&2.8(0)(1)(3)(0)\mbox{\ \ MeV},\\
m_{s}/\hat{m}&=&27.4(1)(4)(0)(1)\nonumber,
\end{eqnarray}
where the errors are from statistics, simulations, perturbation theory and
electromagnetic effects, respectively.

The light hadron spectrum reported in~\cite{MILCNEW} is also an update 
of Fig.~\ref{fig:Nf3GP}.
The systematic errors from chiral extrapolation are not
included in Fig.~16 of~\cite{MILCNEW}, but estimated in
Fig.~4  of~\cite{MILCNEW} for the nucleon mass. 

An intersting work is discussion of the candidate of two-particle state 
for $0^{++}(a_0)$ and $1^{+-}(b_1)$ channel with the observation
of the level crossing as decreasing the quark masses, although
further clarification is still needed with smaller
statistical error and smaller quark masses.

The heavy hadron mass splittings are obtained
in~\cite{Davies:2003ik}, where the NRQCD action and Fermilab
actions are used for $b$ and $c$ quarks respectively as shown
in Fig.~\ref{fig:Nf3GP} (and also Fig. 16 of~\cite{MILCNEW}).

A highlight of prediction from lattice QCD is the mass of $B_{c}$ meson
which is not well established experimentally.
The current experimental value is 6.4(4) GeV from CDF~\cite{CDF}
or 5.95(37) GeV from D$0$ preliminaly data~\cite{Dzero}.
Using the MILC configurations $M_{B_{c}}$ has been calculated by 
the HPQCD-FNAL-UKQCD collaboration~\cite{AllisonMBC} applying the same 
technique to the heavy hadron spectrum described above.
They obtained $M_{B_{c}}=6.304(16)$ GeV.  
Comparions with future improved experiments will provide a verification 
of the technique used for the configuration generation.

\section{Chiral extrapolation of hadronic observables}
\label{sec:chiral_extrap}

Consistency of lattice data with the chiral logarithm
has been an important issue since the lattice 2002
conference~\cite{Bernard:2002yk}.
Possible reasons for not finding the logarithm are :
(1) quark masses are still too heavy to apply ChPT, and/or
(2) lattice cutoff effect distorts the chiral logarithm.
Both effects could be important for available simulation
parameters~\cite{ChPTBaer}. Here we discuss recent results 
bearing on this question. 

\subsection{KS-type fermion action and SChPT}

The MILC Collaboration 
use the staggered chiral perturbation theory (SChPT) as
a guide in the chiral and continuum extrapolations~\cite{MILCCHIRAL,MILCNEW}.
SChPT represents the broken 
$SU(4\times 3)_L\times SU(4\times 3)_R$ symmetry, 
and incorporates both the taste symmetry breaking effect and
fourth-root trick.
The one-loop formula for the pseudo-scalar masses and decay
constants are obtained in
\cite{Aubin:2003mg,Aubin:2003uc,Aubin:2003rg}.
Using the data at different lattice spacing and partially 
quenched~\cite{Bernard:1993sv} data sets,
they simultaneously fit $M_{\mathrm{PS}}^2$ and
$f_{\mathrm{PS}}$, and then the chiral and continuum limit
can be taken at the same time. 
They have two lattice spacings, $a\sim 0.125$ fm (coarse lattice) and 
$\sim 0.09$ fm (fine lattice).
On the coarse lattice, five values of the light quark masses
$a\hat{m}'$ corresponding to the range of
$M_{\mathrm{PS,LL}}$ from 0.60~GeV to 0.25~GeV are simulated,
and the strange quark mass $am'_s$ is fixed at a slightly
higher value than nature and the ``$K$-meson'' mass
$M_{\mathrm{PS,LS}}$ ranges from 0.69~GeV to 0.58~GeV. 
On the fine lattice, two values of $a\hat{m}'$'s at a fixed
$am'_s$ are available to date, which correspond to
$M_{\mathrm{PS,LL}}=$ 0.45--0.32 GeV and 
$M_{\mathrm{PS,LS}}=$ 0.60--0.55 GeV
(the prime on mases means that they are the mass of dynamical quarks 
used in the simulations).
In the fitting the partially quenched points are used, which
could help to stabilize the fit~\cite{Sharpe:2000bc}.
In the MILC analysis the valence light quark masses are 
in the range $0.1m'_s$--$m'_s$.

The taste breaking effect appears in the mass splitting
between Goldstone and non-Goldstone pions.
With the AsqTad action the breaking of $O(\alpha^2 a^2)$ is
expected, which is confirmed by measuring the ratio of the
mass-squared splitting of fine to coarse lattice 
$(\Delta M^2)_{\mathrm{fine}}/(\Delta M^2)_{\mathrm{coarse}}
\simeq
(\alpha a^2)_{\mathrm{fine}}/(\alpha a^2)_{\mathrm{coarse}}
= 0.372$.
The splitting also shows the $SO(4)$ symmetry expected from
SChPT~\cite{LEESHARPE}. 

The mass range where the (S)ChPT is valid is not known a
priori. 
To see how large quark masses can be used in the chiral
extrapolation they attempted the fit for different subsets
of the data.
The data subsets are:
\begin{itemize}
\item subset I (94 data points) with 
  $m_{x}+m_{y}\le 0.4  m'_s$ (coarse) and
  $m_{x}+m_{y}\le 0.54 m'_s$ (fine).\vspace*{-1ex}
\item subset II (240 data points) with 
  $m_{x}+m_{y}\le 0.7 m'_s$ (coarse) and
  $m_{x}+m_{y}\le 0.8 m'_s$ (fine).\vspace*{-1ex}
\item subset III (416 data points) with 
  $m_{x}+m_{y}\le 1.10 m'_s$ (coarse) and
  $m_{x}+m_{y}\le 1.14 m'_s$ (fine),
\end{itemize}
where $m_{x(y)}$ means the mass of valence quark.
They found that the NLO formula is not sufficient even if
for the subset I (lightest set), for which the 
heaviest meson mass is around $\sim 0.5$ GeV.
At this value the NNLO contribution is expected to be around 
$\sim 3.5$\%, which is larger than the statistical error
(0.1\%--0.7\%). 
Since the full NNLO terms for SChPT is not known, they
include analytic terms in the fitting ansatz. 
The NNNLO effect is also investigated.
For future analysis full NNLO form for SChPT is desired.

The fit parameters include the effects of the scaling
violation. 
Taste symmetry breaking terms could vary as
$O(\alpha^2_{s}a^2)$, and 
other parameters include dependence on $O(\alpha_{s}a^2)$ or
$O(\alpha^2_{s}a^2)$ in the global fitting.
Thus, they investigated the following combination of fit
ansatz and data sets to estimate various systematic errors. 
\begin{itemize}
\item NNLO fit on subset I.\vspace*{-1ex}
\item NNLO fit on subset II.\vspace*{-1ex}
\item NNNLO fit on subset III.\vspace*{-1ex}
\end{itemize}
where NNNLO fit is used to interpolate strange quark mass.

\renewcommand{\figscale}{0.4}
\begin{figure*}[tb]
  \centering
  \includegraphics[scale=\figscale,clip]{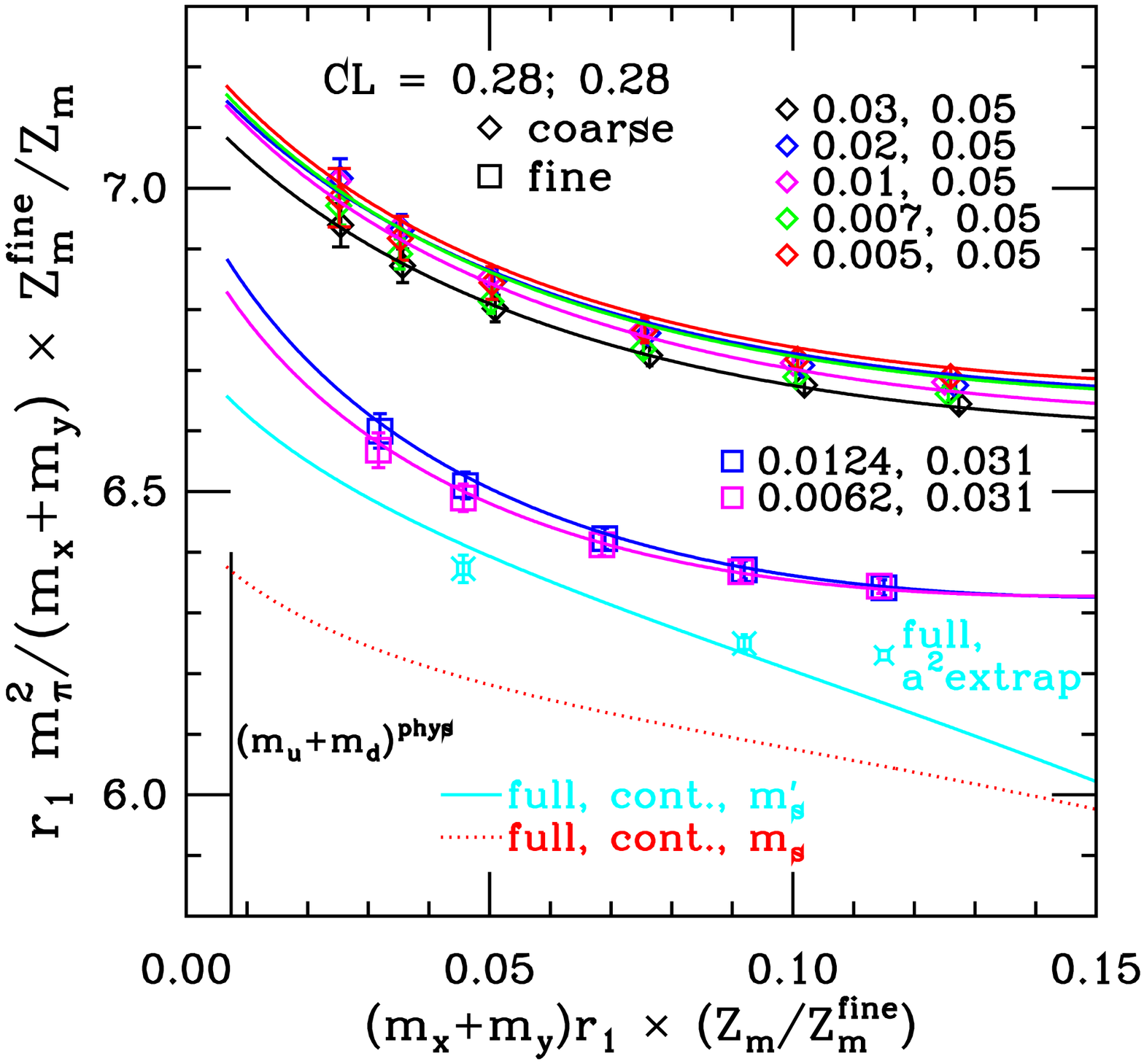}
  \hfill
  \includegraphics[scale=\figscale,clip]{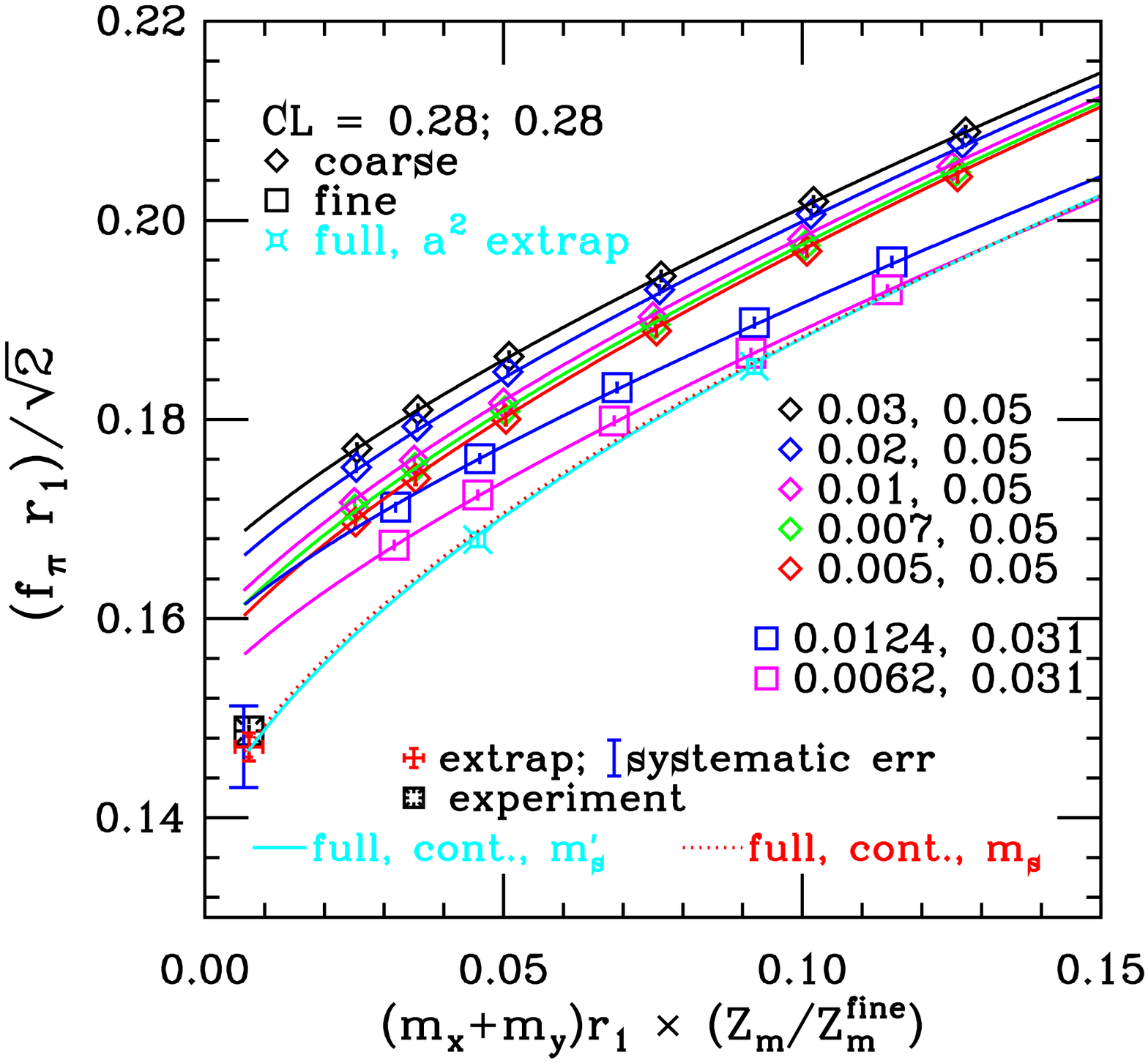}
\vspace*{-2em}
  \caption{
    Global fit with partially quenched data sets from 
    the MILC collaboration~\cite{MILCCHIRAL}. 
    Valence quark mass dependence of
    $m_{\pi}^2/(m_{x}+m_{y})$ (left panel) and 
    $f_{\pi}/\sqrt{2}$ (right panel) in unit of $r_{1}$.
  }
  \label{fig:MILCChPTFit}
\end{figure*}

Figure~\ref{fig:MILCChPTFit} shows the result of
simultaneous fitting (NNNLO fit on subset III).
Selected data points and fit curves are presented.
After the global fitting, the parameters are extrapolated to 
their continuum values, then the ChPT formula at continuum
limit is recovered.
The infinite volume limit is also taken.

Existence of the chiral logarithm could also be examined.
Both the continuum logarithmic terms and taste breaking
logarithms are needed to get good fits.
Although the finite volume effect is expected to be small
with their physical volume ($L>2.5$ fm), 
it is observed that the fit degrades when the finite volume
correction is switched off.
The good fit is highly non-trivial and rely on the good
statistics and partial quenching. 
It is not merely a consequence of the large number of fit
parameters (40 parameters for NNLO fit). 

The physical point is then derived by adjusting $\hat{m}'$
and $m_{s}'$ with the ChPT formula at continuum limit to
the corresponding mess masses and decay constants.
They considered the electromagnetic and isospin violating
effect when adjusting the physical point since their
statistical precision are comparable to these effects. 
Thus the low energy constants, decay constants, quark masses
are obtained~\cite{MILCNEW,Aubin:2004ck}.

They observed the chiral logarithm and obtained the best
results in the chiral/continuum limit.
For further improvement, study on scaling violation using
three or more lattice spacing is desired, with which
the assumptions on the scaling of the coefficients in the ChPT formula
can be verified.
The partially quenched NNLO formula is now available in the
continuum theory~\cite{PQChPTTwo}, which would give hints 
to the NNLO SChPT calculations.

They observed that the ChPT formula gives a natural
expansion with respect to the quark mass and taste symmetry
breaking terms ($O(1)$ coefficients in the expansion).
By a naive order counting NNLO contribution is $\sim$ 10\% even at
$M_{\mathrm{PS}}\sim 600$ MeV.
If we allow 10\% statistical error, I think that NLO formula
can only apply to the data set with
$M_{\mathrm{PS}} \lsim 600$~MeV, 
although MILC's analysis poses more tight restriction.

\subsection{Wilson-type fermion action}

The CP-PACS collaboration extended their coarse
($a=0.22$~fm) lattice simulation~\cite{AliKhan:2000mw} 
toward smaller quark mass region~\cite{Namekawa:2004bi}. 
The simulation parameters are tabulated in
Table~\ref{tab:DYNSIM}. 
The lightest quark mass reaches 
$M_{\mathrm{V}}/M_{\mathrm{PS}}\sim 0.35$.
They investigated the chiral behavior of pseudo-scalar meson
mass, decay constant and PCAC quark mass. 

The chiral fit is carried out with the continuum ChPT
formula as well as the Wilson ChPT
(WChPT)~\cite{Aoki:2003yv} formula, which contains the
effect of scaling violation of $O(a)$ and $O(a^2)$. 
The former contribution can change the coefficient of the
chiral logarithm; 
the latter introduces more singular logarithmic behavior and
important to realize the Aoki phase~\cite{Sharpe:1998xm}.
Since the NLO calculation is not available for the partially
quenched WChPT with the $O(a^2)$ effect, 
the fit is done on the unitary data set.
Figure~\ref{fig:WChPT} shows the fits of the pseudo-scalar
meson mass using the NLO ChPT with and without the scaling
violation effects.
The continuum NLO ChPT can fit data only below 
$M_{\mathrm{PS}}\simeq630$ MeV (top panel), while
the NLO WChPT can fit data up to $M_{\mathrm{PS}}\sim$
1~GeV. 

The crucial test of the WChPT is to see whether it can
explain the lattice spacing dependence of the data, which
has been investigated using the $N_f=2$ data from CP-PACS
at four lattice spacings~\cite{Aoki:2003yv}.
From the fits of available data at heavier mass region the
scaling violation is not well described by the WChPT formula.
They need simultaneous fit on PCAC quark masses, decay constants 
and meson masses,
and smaller quark mass data including partially quenched
data at finner lattice spacings in order to identify the
$O(a)$ and $O(a^2)$ effects in the chiral fits.

\renewcommand{\figscale}{0.57}
\begin{figure}[tb]
  \centering
  \includegraphics[scale=\figscale,clip]{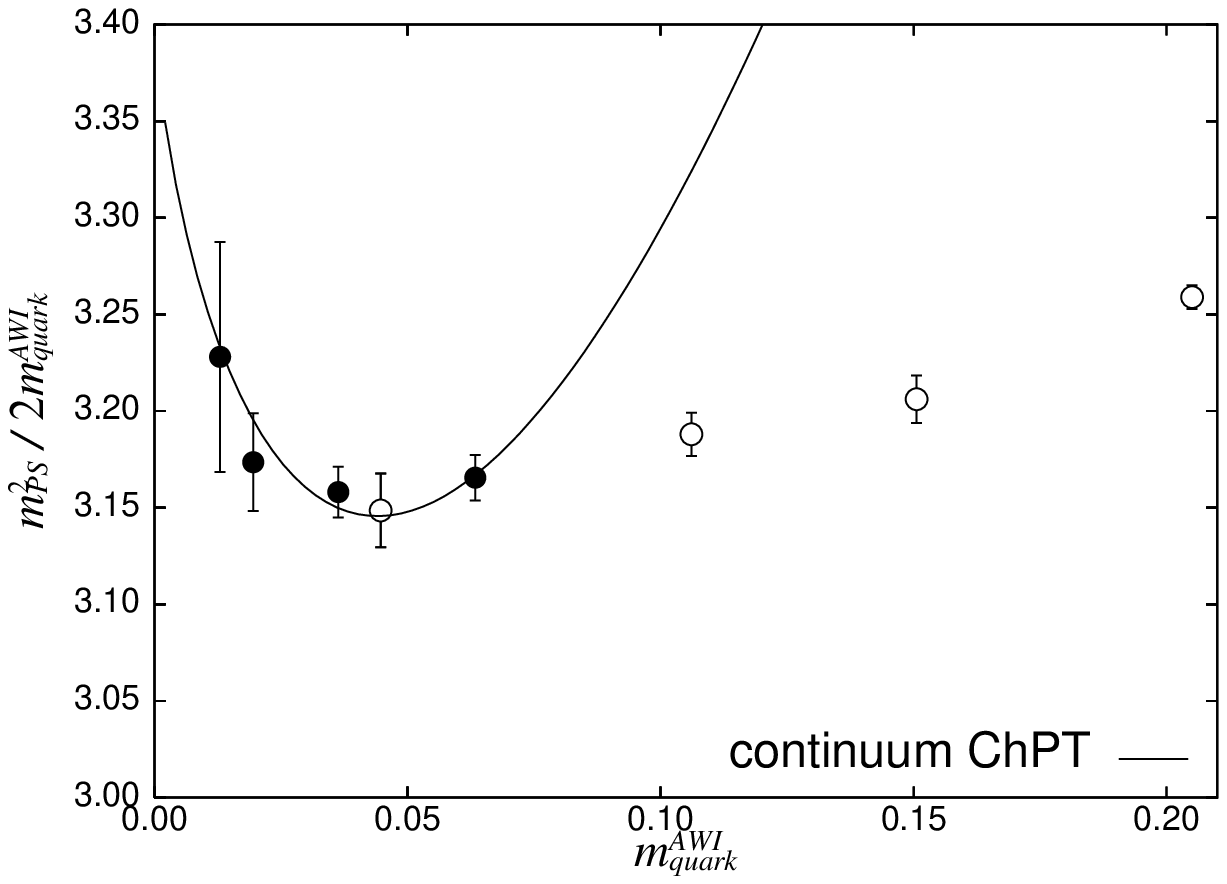}\hfill
  \includegraphics[scale=\figscale,clip]{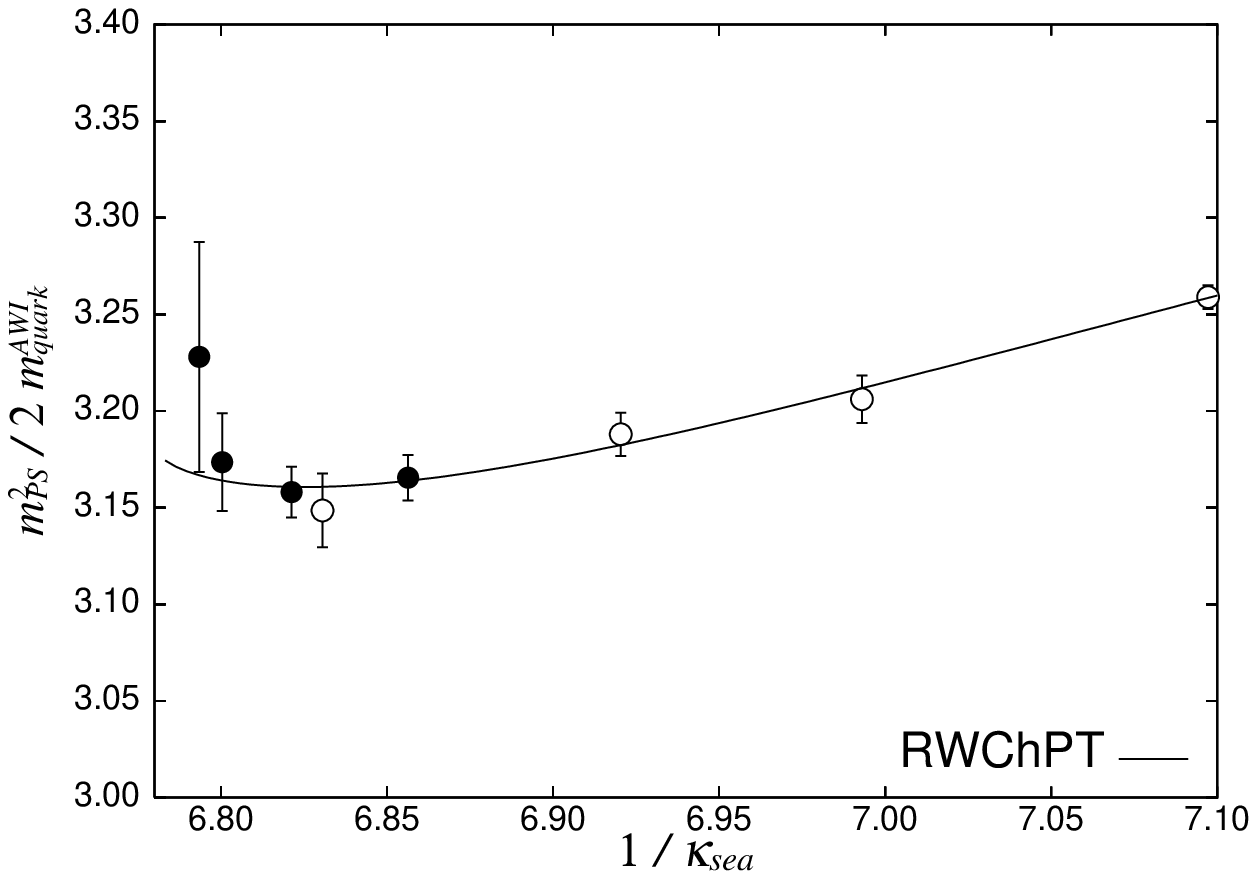}
\vspace*{-2em}
  \caption{
    The continuum ChPT fit to the data 
    $M_{\mathrm{PS}}/M_{\mathrm{V}}<0.6$
    (top panel), and the resummed Wilson ChPT fit to the
    whole data (bottom panel)~\cite{Namekawa:2004bi}.
  } 
\vspace*{-2em}
  \label{fig:WChPT}
\end{figure}

The qq+q collaboration studied the chiral fit with the
unimproved Wilson
action~\cite{Farchioni:2003nf,Farchioni:2004tv} using
the NLO WChPT formula including the $O(a)$
effects~\cite{Rupak:2002sm}.
The simulation parameters are shown in Table~\ref{tab:DYNSIM},
which cover the similar region as in \cite{Namekawa:2004bi}.
They also estimate the (continuum) analytic NNLO
contribution and the low energy constants of ChPT.

Using the double ratio method~\cite{Hashimoto:2002vi} they
estimate the contribution of NNLO and $O(a)$ effect. 
Their data indicate that the $O(a)$ contribution is not important,
but the NNLO is. 
It seems that this contradicts with the naive order counting
for $a\sim$ 0.2~fm (see also \cite{ChPTBaer}).
Because the qq+q data point $\beta=5.1$ is close to the
recently observed first order phase transition point
$\beta=5.2$~\cite{Farchioni:2004ma,Farchioni:2004us}, and
one may suspect that the data could be badly distorted.
Therefore, one needs a detailed study of scaling to draw
definite conclusion.

\renewcommand{\figscale}{0.60}
\begin{figure*}[tb]
  \centering
  \includegraphics[scale=\figscale,clip]{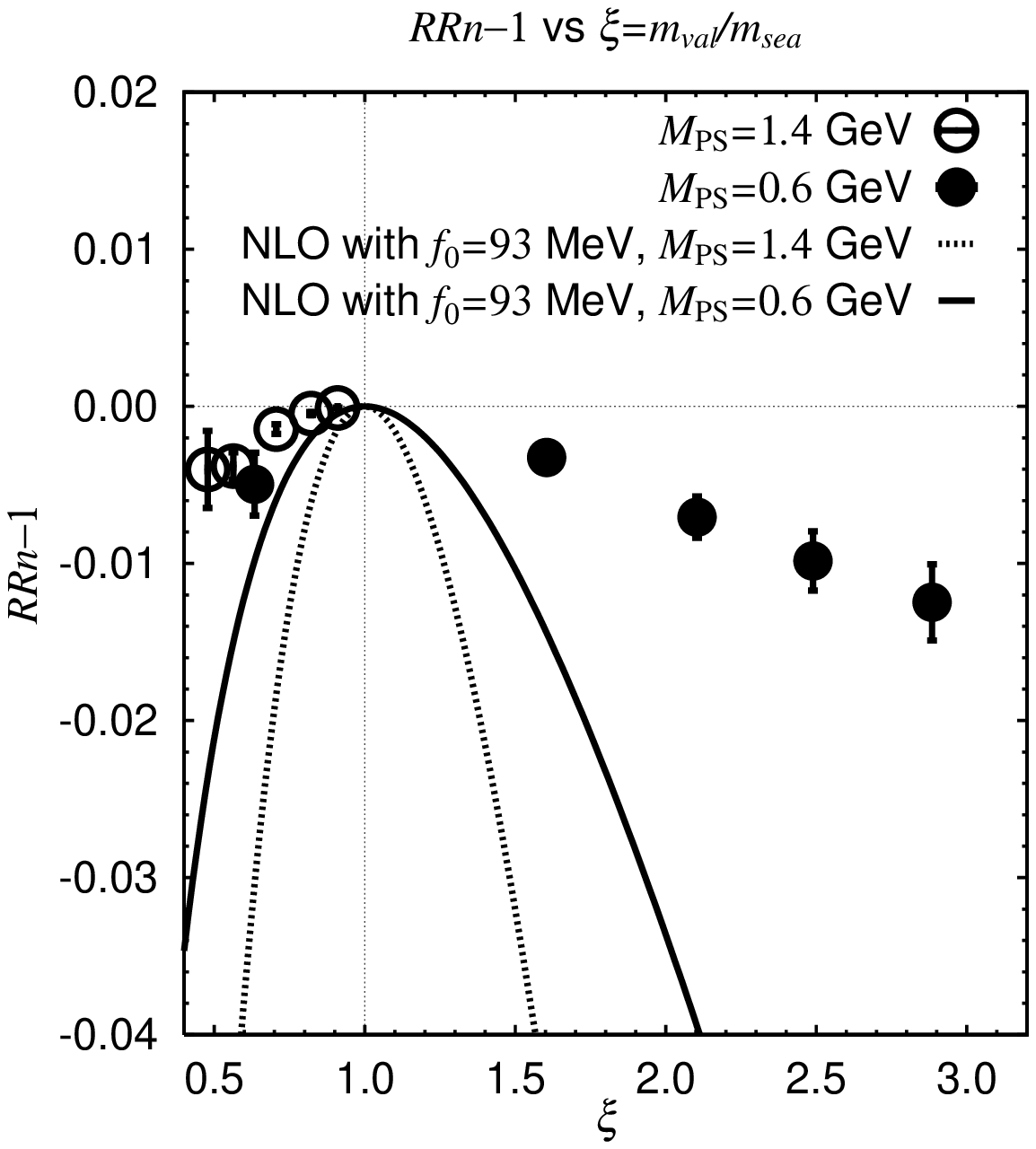}\hfill
  \includegraphics[scale=\figscale,clip]{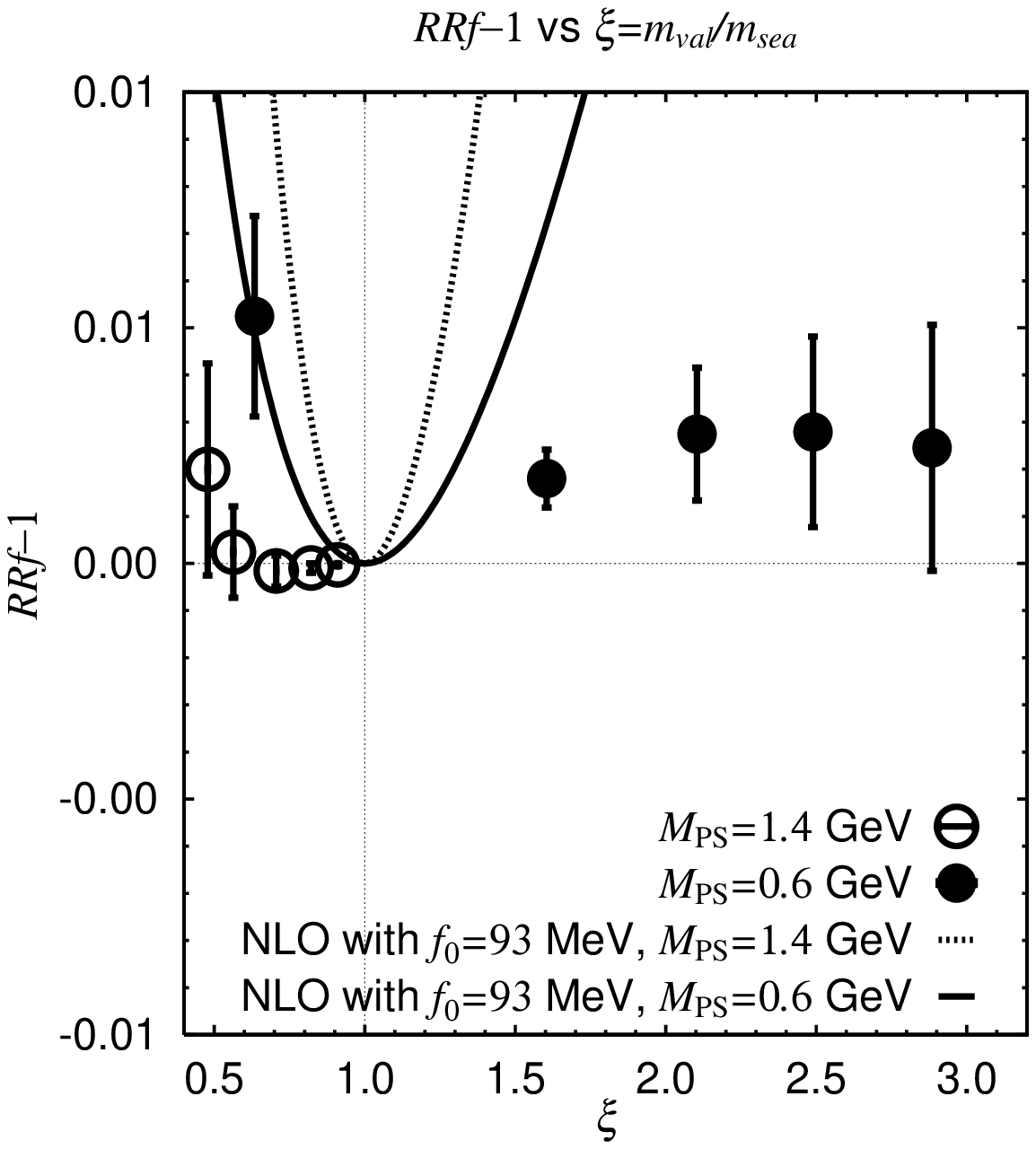}
\vspace*{-2em}
  \caption{
    $RRn$ and $RRf$ as a function of $\xi$, data from JLQCD,
    $20^{3}\times 48$,  
    $M_{\mathrm{PS}}=$0.6--1.4 GeV~\cite{Aoki:2002uc}
  }
\vspace*{-0em}
  \label{fig:JLDBL}
\end{figure*}

To further discuss this point, we show 
in Figure~\ref{fig:JLDBL} the double ratio against the
ratio of valence and sea quark masses $\xi=m_{V}/m_{S}$ from
the partially quenched data at $a\sim$ 0.09~fm of the JLQCD 
collaboration~\cite{Aoki:2002uc,SHJLPQ}.
$RRf$ and $RRn$ are defined as
\begin{eqnarray}
  RRf & = & \frac{f_{VS}^2}{f_{SS}f_{VV}}, 
  \\
  RRn & = & \frac{4\xi}{(1+\xi)^2}
  \frac{M_{VS}^4}{M_{SS}^2M_{VV}^2},
\end{eqnarray}
with $V$ denoting $m_{\mathrm{val}}\ne m_{\mathrm{sea}}$ and
$S$ means the unitary point~\cite{Farchioni:2003nf}.
The lines are the theoretically expected chiral logarithm at
NLO in the continuum ChPT. 
Filled symbols ($M_SS\simeq$ 600~MeV) below $\xi=1$ are
close to the corresponding theoretical expectation, compared
to the heavier quark mass data. 
It suggests that the ChPT is valid only below 
$M_{\mathrm{PS}}<0.6$ GeV.
The detailed scaling study and application of partially
quenched WChPT formula including $O(a^2)$ effect is,
however, needed again.

The UKQCD collaboration added a new data at
$M_{\mathrm{PS}}/M_{\mathrm{V}}\sim 0.44$  
($M_{\mathrm{PS}}\sim 400$ MeV)~\cite{Allton:2004qq} to 
the previous simulations~\cite{Allton:1998gi}. 
Large finite volume effect is expected, since their physical
volume is $L\sim$ 1.5~fm and $M_{\mathrm{PS}}L=3.2$ at the
lightest data. 
They observed some curvature for the pseudo-scalar meson
decay constant as a function of pseudo-scalar meson mass
squared as shown in Figure~\ref{fig:UKfPI}. 
The bursts are the experimental values of $f_{\pi}$ and $f_{K}$
and the solid curve is the continuum ChPT curve fitted to
the experimental values. 
The open boxes are the lattice data.
The crosses are the decay constants in the finite volume
($L=$ 1.5~fm) expected from continuum ChPT
formula~\cite{Colangelo:2003hf}. 
They argue that the curvature observed in lattice data is
the result of finite volume effect from the chiral
logarithm. 
Further clarification is, however, needed through the
systematic study on the volume dependence and quark mass
dependence. 

\renewcommand{\figscale}{0.38}
\begin{figure}[tb]
  \centering
  \includegraphics[scale=\figscale,clip]{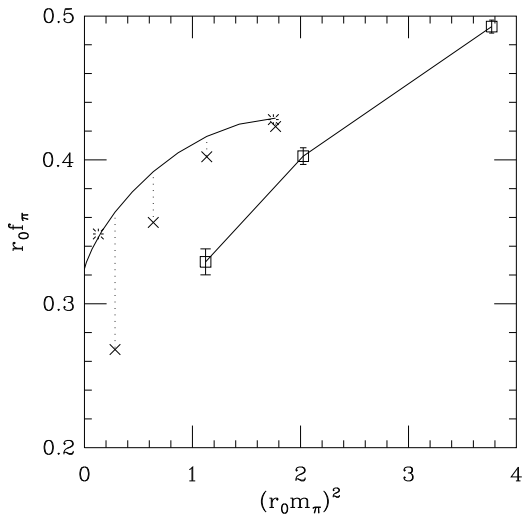}
\vspace*{-3em}
  \caption{
    The pseudo-scalar decay constant as a function of the pseudo-scalar
    meson mass squared from UKQCD~\cite{Allton:2004qq}.
  }
\vspace*{-2em}
  \label{fig:UKfPI}
\end{figure}

\subsection{twisted mass QCD and a surprise of a first-order phase transition}

Since the chirally twisted mass~\cite{Frezzotti} introduces a
lower bound on the eigenvalue of the (hermitian)
Wilson-Dirac operator, the unquenched simulation could be
substantially faster for small quark masses compared to the
(untwisted) Wilson-type quarks.
Two-flavor twisted mass Wilson quark simulations have been
started and the first results are reported
by~\cite{Farchioni:2004ma}.
Surprisingly, they found a strong metastability in the
plaquette expectation value at $\beta=5.2$ for twisted mass
of $\mu=$ 0--0.1.
The quark mass determined through the PCAC relation changes
its sign at the phase gap, and the pion mass does not vanish
by tuning hopping parameter.
The origin of the metastability may be attributed the
possible scenario described in~\cite{Sharpe:1998xm}, 
that is that the negative coefficient of the $O(a^2)$ term
in the chiral effective Lagrangian could cause the absence
of the Aoki phase. 
This observation raises a new question on the phase
structure in the $(\beta, \kappa)$ plane in unquenched 
simulations with Wilson type quarks.
To reach the chiral limit at finite lattice spacings one
needs the Aoki phase separated by a second order phase
transition, but it is not necessarily the case.
Then, the chiral extrapolation must be done after taking the
continuum limit.
The extension of the phase diagram to twisted mass direction
is also investigated in the effective
Lagrangian~\cite{Sharpe:2004ps}. 

\subsection{Domain-wall fermion}

The RBC collaboration performed two-flavor dynamical
simulations with the domain-wall quark
action~\cite{Izubuchi:2003rp,RBCNEW}. 
It is expected that the Ginsparg-Wilson (GW) type fermion has much better
chiral property because of the exact chiral symmetry at
finite lattice spacings.
The domain wall fermion, however, in finite extent in the fifth 
dimension does not obey the GW relation exactly. 
The violation of the symmetry can be measured by the residual mass 
defined through the Ward-Takahashi identity of the DW fermion and
the residual mass remains even in the massless limit.
The lattice size in the fifth dimension is $N_s=12$, with
which the residual mass is small enough
$am_{\mathrm{res}}\simeq0.001$.  
Using three quark masses 
$m_{\mathrm{sea}}\sim m_{s}/2$, $3m_s/4$, $m_s$,
they analyzed the pseudo-scalar meson mass and the decay
constant using continuum NLO ChPT formula.
They observed that for the pseudo-scalar meson mass
the partially quenched NLO ChPT can fit the data for
$\le 3 m_{s}/4$ ($M_{\mathrm{PS}}\lsim 630$ MeV),
while $\chi^2$ gets much worse if one includes the heavier
mass data ($M_{\mathrm{PS}}\sim$ 690~MeV).
For the decay constant, the NLO formula cannot describe the
data well even if one restrict the masses in $\le 3 m_{s}/4$ 
($M_{\mathrm{PS}}\lsim 630$ MeV).
This is probably because of limited available data points
compared to that of meson mass case. 
Although they observed 
$f_{\pi}/m_{\rho}=0.170(8)$, and $f_{K}/m_{\pi}=1.18(1)$
from the LO fit in fair agreement with experiment, the full
NLO analysis with more data point would be needed to control
the chiral extrapolation.

\subsection{Summary}

For the KS-type fermions 
the chiral logarithm is observed by the MILC collaboration
for small sea quark masses using the SChPT formula. 
On the other hand, there is no definite conclusion for the Wilson-type 
fermions primarily because the unquenched simulations are still limited 
to relatively heavy quark masses. As a result, 
chiral extrapolation with appropriate ChPT formula in
finite lattice spacing is not well investigated so far.
Nevertheless, from the observation from the MILC results and 
the studies discussed in this section, it is likely that 
ChPT can only be applied at $M_{\mathrm{PS}} \lsim 0.6$ GeV. 
For Wilson-type fermions, development of simulation algorithms 
that allow to enter this region~\cite{Kennedy} is crucial for 
progress.

\section{Other topics on hadron spectrum}
\subsection{Mixed action simulations}
The Ginsparg-Wilson fermions improve the chiral property. 
However, it is numerically so demanding that the unquenched
simulations are hard task especially for light quarks.
Use of different quark actions for valence and sea quarks
can provide a clue to avoid too large computational cost
while partly keeping the good chiral property.
The partially quenched quark mass combinations can also be
studied in this setup.
A possible problem is the violation of unitarity, which can
be removed only in the continuum limit.

As I already mentioned, the unquenched configurations with
KS-type sea quarks reaches the chiral regime. 
Various measurements on these configurations have been carried
out. 
The light hadron spectrum in Section~\ref{subsec:KSLH} may 
also be considered as a mixed action calculation because the
treatment of the KS-Dirac operator is different between
valence and sea; while sea quarks are expressed by the
fourth-root trick, valence quarks are treated by the quantum number
projection.

Preliminary results for the light hadron spectrum with
the overlap or domain wall valence quarks on the $N_f=2+1$
KS-type sea quarks generated by the MILC collaboration have been 
reported in \cite{Schroers,Tweedie}. 
The valence mass is tuned by matching pseudo-scalar meson
masses calculated with the domain wall and KS quarks~\cite{Schroers}. 

\subsection{Flavor singlet meson mass}
The large splitting between the flavor singlet $\eta'$ mass and 
other flavor octet meson masses should be 
explained by lattice QCD simulations. 
There have been a series of studies calculating the flavor
singlet mass in unquenched simulations~\cite{Struckmann:2000bt,McNeile:2000hf,Lesk:2002gd,Schilling:2004kg,Allton:2004qq,Venkataraman:1997xi}.

\renewcommand{\figscale}{0.35}
\begin{figure}[tb]
  \centering
  \includegraphics[scale=\figscale,clip]{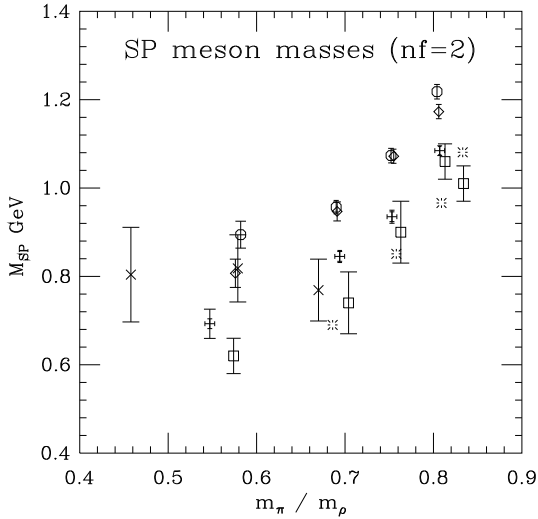}
  \vspace*{-4em}
  \caption{
    The $\eta'$ meson mass in $N_f=2$ as a function of 
    $M_{\mathrm{PS}}/M_{\mathrm{V}}$ quoted from~\cite{Allton:2004qq}.
    Results are obtained with 
    PC~\cite{Allton:2004qq,McNeile:2000hf} (crosses),
    PW~\cite{Schilling:2004kg} (bursts),
    PW~\cite{Struckmann:2000bt} (squares),
    and 
    RC(TP)~\cite{Lesk:2002gd} (diamond ($\beta=2.1$), 
    octagon ($\beta=1.95$), fancy plus ($\beta=1.8$))
    actions.
  }
\vspace*{-2em}
  \label{fig:ISOSINGLET}
\end{figure}

The continuum limit of the $\eta'$ mass in $N_f=2$ was
studied for the RC(T) action~\cite{Lesk:2002gd}, 
and $M_{\eta'}=960(87)(^{+\ 36}_{-248})$ MeV was obtained
where the systematic errors come from the continuum and 
chiral extrapolations. 
This value seems consistent with the
experimental value $M_{\eta'}=$ 956~MeV.  
However, the lattice result represents the value in the $N_f=2$ world 
for which the expectation for the iso-singlet meson mass is 
861~MeV (method described
in~\cite{Allton:2004qq,McNeile:2000hf}) 
or 715~MeV (from Witten-Veneziano formula).
The above result from the RC(T) action is consistent within
the error. 
The rather large systematic error in the negative direction
comes from continuum extrapolation. 
It seems difficult to obtain a reliable continuum
extrapolation because of large statistical errors even
if we add some points with different lattice spacings around
$a^{-1}=$ 2.5~GeV.
One would need a better method to improve the statistical
signal. 
The chiral extrapolation should be carried out for both
$M_{\mathrm{PS}}^2$ and $M_{\mathrm{PS}}$ as a linear function of 
$M_{\mathrm{PS}}^2$, since reliable fit ansatz is not known.

Figure~\ref{fig:ISOSINGLET} shows the
$M_{\mathrm{PS}}/M_{\mathrm{V}}$ dependence 
of $M_{\eta'}$ compiled in \cite{Allton:2004qq}. 
The data show a similar dependence on $M_{\mathrm{PS}}/M_{\mathrm{V}}$ 
except for crosses.
The data at lighter two crosses~\cite{Allton:2004qq} ($a=0.1$ fm) 
seems to be constant. 
Squares~\cite{Struckmann:2000bt} are obtained 
with $Z_{2}$ noise method, while
bursts~\cite{Schilling:2004kg,Neff:2004kf} are obtained with
truncated eigenvalue approximation (TEA) method from the
same SESAM configuration~\cite{Eicker:1998sy} 
(symbols are slightly shifted horizontally to avoid
overlapping). 
They are consistent with each other while the TEA has
smaller statistical error. 
The value at physical point is different:
$M_{\eta'}\sim 290$ MeV with 
TEA~\cite{Schilling:2004kg,Neff:2004kf} and
$\sim 520^{+125}_{-58}$ MeV with $Z_2$~\cite{Struckmann:2000bt}.
Systematic study of chiral extrapolation with more data at
smaller quark masses will clarify the difference of these data. 
The 10--20\% scaling violantion from
$O(a\Lambda_{\mathrm{QCD}})$ error is also expected with unimproved 
gauge and quark actions even at $a^{-1}\sim 2.3$ GeV lattice.

The effect of $\eta$ and $\eta'$ mixing can be treated 
by the partially quenched strange analysis within the $N_f=2$ 
simulations~\cite{Schilling:2004kg,Neff:2004kf,McNeile:2000hf,Venkataraman:1997xi}.
The quadratic mass matrix in the quark-flavor basis is 
obtained in~\cite{Schilling:2004kg,Neff:2004kf,McNeile:2000hf}
on the $a^{-1}\sim 2.3$ GeV SESAM configurations and
they obtained $M_{\eta}=292(31)$ MeV and $M_{\eta'}=686(31)$ MeV
after the diagonarilzation of the mass matrix. 
While these individual numbers are still below the
experimental numbers, the splitting is comparable to the
experimental value. 

\subsection{Universality of quenched hadron spectrum}
Sometime ago, it was pointed out that the values of 
$M_{\mathrm{N}}/M_{\mathrm{V}}$
at $M_{\mathrm{PS}}/M_{\mathrm{V}}=0.5$ disagree
in the continuum limit between the PKS action and PW
(RC(TP)) action~\cite{Aoki:2000kp}. 
The PKS data in $aM_{\mathrm{V}}<1.4$ were extrapolated with
a function $c_0 + c_1\times a^2$, and the PW (RC(TP)) data in
$aM_{\mathrm{V}} <1$ were extrapolated with $c_0+c_1 \times a$.

This year reanalysis is done by adding new data with various
lattice actions generated since 2000~\cite{Davies:2004hc}. 
The universality is investigated in the quenched
approximation by checking the consistency in the continuum
limit among the various discretization method using the
Bayesian fitting approach~\cite{Lepage:2001ym}. 
All the data are simultaneously fitted with some constraints.
The fit functions are polynomials which contains leading and
higher order terms as a function of $aM_{\mathrm{V}}$ 
(for the KS-type action only $a^{2n}$ terms are included) with
a common continuum limit.
The coefficients are constrained to stay within reasonable
values by the Bayesian prior. 
Thus, they obtained a reasonable $\chi^2$ value for the
consistent continuum limit.
The continuum value is $2\sigma$ below 
and $4\sigma$ below from the previous analysis of the PK and
PW actions respectively.
They also investigated the universal continuum limit
of $M_{\mathrm{V}}r_1$ and $M_{\mathrm{N}}r_1$ with more 
data (lattice actions) and obtained a reasonable $\chi^2$ values.
These tests suggest the difficulty to estimate the
systematic error of continuum extrapolation from single
lattice action and the importance of the action improvement
with many lattice spacings.  
The similar test for universality should eventually be done
for unquenched simulations.

\section{Conclusions}

The unquenched QCD simulations with dynamical up, down and strange quarks 
have made significant development over the past year.  The progress has 
been more notable for the KS-type fermions for which an extensive work 
has been made both in pushing 
the simulations toward small quark masses and in analyzing the results. 
Applying staggered ChPT, it was shown that their data are consistent with 
the expected logarithmic chiral behavior once the pseudo scalar meson 
masses are decreased below $M_{\mathrm{PS}}<0.6$ GeV.  
Worries, however, regarding the field theoretic foundation of 
the fourth-root trick mandatory in $N_f=2+1$ simulations with 
the KS-type fermions still need to be clarified.  

Simulations with the Wilson-type fermions yielded an encouraging result 
that the meson spectrum moves progressively close to experiment as the number 
of dynamical quarks increases from $N_f=0$ to $2$ to $2+1$.  
The largest issue is the long chiral extrapolation from heavy quark masses involved 
in reaching this result; consistency with the expected chiral behavior is 
not established to satisfaction. Hence algorithms which allow simulations 
with quark masses as light as used in the KS-type fermions are deeply needed.  
As we encountered for nucleon mass in the quenched case, 
cross checks between the KS-type and Wilson-type 
fermions which will be made possible by such algorithms 
are very important for establishing unquenched predictions. 

Looking further ahead, unquenched simulations with the Ginsparg-Wilson 
type quark actions with the correct chiral behavior are the most challenging 
issue in future lattice QCD.

\vspace*{1ex}
I would like to thank all the colleagues who made
their results available before the conference.
I also thank S.~Hashimoto and A~Ukawa for valuable comments 
and proof reading on the manuscript.
This work is supported in part by the Grants-in-Aid
of Ministry of Education (Nos. 16740147, 15204015).

\end{document}